\documentclass[preprint,12pt]{elsarticle}

\usepackage{amssymb}
\usepackage{graphicx}

\def\CC{{\mathbb C}}

\def\RR{{\mathbb R}}

\def\PP{{\mathbb P}}

\def\ee{{\mathrm e}}

\def\ii{{\mathrm{i}}}

\def\cE{{\mathcal E}}

\def\cM{{\mathcal M}}

\def\cZ{{\mathcal Z}}

\def\ri{{\mathrm i}}
\def\rr{{\mathrm r}}

\def\book#1{\rm{#1}, }

\def\jour#1{\rm{#1}, }

\def\vol#1{\textbf{#1}}

\def\by#1{{\rm{#1}, }}

\newtheorem{theorem}{Theorem}[section]

\newtheorem{lemma}[theorem]{Lemma}

\begin{document}

\begin{frontmatter}



\title{An algebro-geometric model for the
shape of supercoiled DNA}


\author{Shigeki Matsutani}
 \address{
Graduate School of Natural Science \& Technology, 
Kanazawa University Kakuma Kanazawa, 920-1192, Japan}
\ead{s-matsutani@se.kanazawa-u.ac.jp}

\author{Emma Previato}
 \address{
Department of Mathematics and Statistics,
Boston University, Boston, MA 02215-2411, U.S.A.}
\ead{ep@math.bu.edu}

\begin{abstract}
This article proposes a  model including
thermal effects for closed supercoiled DNA. 
Existing models include an elastic rod.
Euler's elastica, 
 ideal elastic rods on a plane, have only two kinds of closed
 shapes, the circle and a figure-eight, realized as minima
 of the Euler-Bernoulli energy.
Even  considering three dimensional effects, 
this elastica model
provides much simpler shapes than  observed  via
 Atomic-Force Microscope (AFM),
since the minimal points of the energy 
are expressed by  elliptic functions. 
In this paper, by a generalization of elastica, we obtain shapes 
determined by data of hyperelliptic curves, which partially 
reproduce the shapes and properties of the DNA. 
\end{abstract}



\end{frontmatter}



\section{Introduction}
Using the atomic-force microscope (AFM) and
the electron microscope (EMS), 
certain configurations of the 
supercoiled DNA (deoxyribonucleic acid), especially plasmid DNA, have been
observed \cite{LS}.
The shapes of the supercoiled DNA had previously been  studied 
\cite{CDLT, KB, VV}.
The shapes show that the large-scale
structure of DNA might have elasticity.
The elastic rod model of the supercoiled DNA was proposed and
investigated  \cite{BM, SH, TsuruWadati, WadatiTsuru}.

A rod with elasticity gives rise to an ``elastica''; these 
 were studied by
Euler and the Bernoullis \cite{Mat10, Truesdell83}. 
They showed that the shapes of the 
elastica on a plane are realized as 
the minima of the Euler-Bernoulli (EB) energy
and are described by  elliptic functions. 
Mathematically, they considered a class of
analytic immersions $Z: [0,1] \to \CC$ of curves in the
complex plane $\CC$.
Daniel Bernoulli produced the elastica
by  the minimal principle 
with respect to the EB energy $\displaystyle{
\cE_{\rm{EB}}[Z]:=\oint \kappa^2(s) d s}$ where $\kappa(s)$ is the curvature, 
\begin{equation}
\kappa(s):=\partial_s \phi(s), \quad
\phi(s):= \frac{1}{\ii}\log \partial_s Z(s),
\label{eq:kappa}
\end{equation}
and $s$ the arclength  of the rod. In \cite{Euler44},
Euler showed that under the isometric constraint, the shape
is expressed by  elliptic integrals
and used elliptic integrals to find the elastica trajectory.
Euler  classified and listed the shapes of the elastica
 \cite{Euler44, Love, Mat10}.
He showed that the looped elastica
 on the plane are only of two types, the circle and the figure-eight. 
Even in  three dimensional space,
the elastica, which were studied by Kirchhoff and Born \cite{Born}, 
are also
expressed by  elliptic functions \cite{SH, Shin, Tsuru}.

Since the elliptic function is related to the compact Riemann surface of
genus one and has only double periods, 
the shape expressed by the elliptic functions cannot exhibit the 
complicate supercoils
 observed in DNA \cite{GPL, MPB, SH, St, Tsuru, TsuruWadati, WadatiTsuru}.
Therefore, although  the supercoiled DNA has been 
much studied,
 the  shape pertaining to the elastic-rod model is limited to Euler's list
\cite{GPL, MPB}.

It is a question why the supercoiled DNA observed by AFM
has 
shapes more complicated than the ones in Euler's list.
Several reasons are proposed, e.g.,
 chemical effects (electrostatically charged model in fluid) \cite{LKS}.
The thermal effect is also important:
indeed, in 
\cite{KP, Petal}, the shape of the supercoiled DNA
is derived 
by considering the thermal effect on
molecular dynamics.

In order to derive the shape of the supercoiled DNA,
we proposed the statistical mechanics of elastica,
which allows for  the thermal effect,
as a generalized elastica problem (sometimes called quantized elastica)
\cite{Mat97}
and studied the corresponding dynamics in
 \cite{Mat07, 
Mat08,Mat10,MO03a,MP16}.
In other words, we consider the partition function
$\displaystyle{
\cZ[\beta]= \int_{\cM} DZ \exp(-\beta \cE_{\rm{EB}}[Z])}$,
where 
$
\cM
:=\{ Z: S^1 \hookrightarrow \CC\ | \ \mbox{analytic, isometric} \}$.
We classify the moduli space $\cM$ according to the energy 
$\cE_{\rm{EB}}[Z]$.
 The isometric and iso-energy deformation of 
$Z$ is determined so that the curvature 
$q:=\kappa/2$ obeys the modified Korteweg-de Vries (MKdV)
equation \cite{GoldsteinPetrich1,Mat97, MO03a, MP16, P},
\begin{equation}
	-\partial_t q + 6 q^2 \partial_s q + \partial_s^3 q
	 =0, 
\label{4eq:MKdV}
\end{equation}
where $\partial_t :=\partial/\partial t$ and $\partial_s :=
\partial/\partial s$, which is also known as the loop soliton
equation. 
We briefly review this setting in Section \ref{sec:SME}, which provides
the mathematical-physics foundation for the equation.
We showed that 
the shapes of the elastica  in the iso-energy classes of the 
excited states of the EB
energy are determined by  hyperelliptic functions of higher 
genus \cite{Mat02b, MP16}.
Since the hyperelliptic functions are a generalization of
elliptic functions, our results are a natural extension of the 
Euler-Bernoulli theory of elastica \cite{Mat10}.
Moreover, they can be  extended to  three dimensional
space \cite{GoldsteinLanger,Mat99a}: the MKdV hierarchy is replaced with
the nonlinear Schr\"odinger (NLS) and complex mKdV (CmKdV) hierarchies.
In fact, the hyperelliptic solutions of the generalized elastica,
 can be extended to three-dimensional space, in view of work on the nonlinear
 Schr\"odinger and complex mKdV hierarchies \cite{EEK, P0, Shin}. We do not
 present them in this paper, but at the end of subsection \ref{sec:g=2}
we propose them as model for DNA.  

However, abstract hyperelliptic function theory 
is not amenable to computation, and to the derivation of
explicit shapes.
We developed  Abelian function theory, including  hyperelliptic
\cite{Mat02b,Mat07, Mat08, Mat10, MO03a, MP16}, 
by replacing theta functions with  sigma functions
\cite{Baker97, BEL97b}, and those results now allow us 
to address the question of  supercoiled DNA.

In this paper, we argue the possibility of
the realization of  the shapes 
in terms of  hyperelliptic functions of genus two, 
based on the solutions found in \cite{Mat02b, Mat07, Mat08}.
Although theoretically we need
higher-genus $g$ hyperelliptic curves ($g>2$)
to construct the solution of (\ref{4eq:MKdV}), 
still
we obtain closed orbits using 
genus-2 curves and
argue that they 
 reproduce the properties of the supercoiled DNA.

In conclusion, we propose a novel algebro-geometric
investigation of the supercoiled DNA.

\section{Statistical mechanics of elastica}
\label{sec:SME}

In this section, we review the statistical mechanics of elastica
in the framework of mathematical physics.
In statistical mechanics, the investigation of toy models
is  important, even though they are not directly related to 
physical systems \cite{Baxter,ID, Thom}.
The statistical mechanics of elastica was proposed as one of them 
\cite{Mat97, Mat99a}. It is intended to model the shapes of the 
supercoiled DNA but for simplicity we start on the plane.
Observations of supercoiled DNA by AFM and EMS
 show  distorted figure-eight shapes and
circles, cf., e.g.,  \cite{LS}.
Frequently, models of statistical mechanics are connected with
nonlinear equations: our model is  related to
the
modified Korteweg-de Vries (MKdV)
hierarchy.

As mentioned in the Introduction,
we consider the partition functions of elastica,
\begin{equation}
\cZ[\beta]= \int_{\cM} D Z\, \exp(-\beta \cE_{\mathrm{total}}[Z]),
\label{eq:cZ}
\end{equation}
where $\cM$ is the configuration space of the elastica
modulo  the action
of a subgroup of the group of Euclidean motions, i.e.,
$\cM
:=\{ Z: S^1 \hookrightarrow \CC\ | \ \mbox{analytic, isometric} \}/ \sim$
and $D Z$ is a kind of Feynman measure, which is physically defined
but not mathematically rigorous.
Let $L$ be the length of the elastica.
In this paper, we basically consider only the EB energy 
$\cE_{\rm{EB}}$,
\begin{equation}
\cE_{\rm{EB}}[Z]:=\oint \kappa^2(s) d s,
\end{equation} 
for the curvature  $\kappa(s)$ of $Z$  in (\ref{eq:kappa})
and the arclength $s$,
but we could assume that 
the energy $\cE_{\mathrm{total}}[Z]$ in (\ref{eq:cZ}) 
contains  effects from the number of self-contact points
$i(Z)$ and the winding
number 
$\displaystyle{w(Z):=\frac{1}{2\pi} (\phi(L)- \phi(0))}$
of $Z$, e.g., 
\begin{equation}
\cE_{\mathrm{total}}[Z] = \cE_{\rm{EB}}[Z]  + \alpha_1 i(Z)
+ \alpha_2 w(Z).
\label{eq:cEZ}
\end{equation}
with certain coupling constants $ \alpha_1$ and 
$\alpha_2$. 

We could also take  other effects into account.
By standard consideration
in  statistical mechanics,
the partition function
is reduced to 
\begin{equation}
\cZ[\beta]= \int_0^\infty  dE\,  \mathrm{Vol}(\cM_E) \exp(-\beta E),
\label{eq:cZdE}
\end{equation}
where $\mathrm{Vol}(\cM_E)$ is the density of states,
which is given as the volume of the subspace $\cM_E$,
$$
\cM_E:=\{ Z \in \cM\ |\ \cE[Z] = E\ \}.
$$
In order to define the measure  
in (\ref{eq:cZdE}) rigorously, we require more precise geometrical
and analytical knowledge of the elements of $\cM_E$, which is not yet
available.
One of the purposes of this paper is to identify such elements by explicitly
solving the hierarchy.
The integration in (\ref{eq:cZdE}) therefore means that 
the partition function is evaluated by
the computation of
the volume of the states with energy $E$. 

As in \cite{MO03a, MP16},
we classify the moduli space $\cM$ according to the energy 
$\cE_{\rm{EB}}[Z]$. Then as mentioned in \cite{Mat97, MP16},
the isometric and iso-energy deformation of 
$Z$ is determined so that the curvature 
$q:=\kappa/2$ obeys the MKdV
hierarchy \cite{GoldsteinPetrich1,Mat97, MO03a, MP16, P},
\begin{equation}
\partial_{t_\ell} q = \Omega^\ell \partial_s q,
\label{4eq:MKdVh}
\end{equation}
where $\partial_{t_\ell} :=\partial/\partial t_\ell$ and $\partial_s :=
\partial/\partial s$,
$\Omega:= \partial_s^2 + 4 \partial_s (q \partial_s^{-1} q)$
and $t_\ell$ are infinitely many real-time parameters $\ell = 1, 2, 3, \ldots$.
When $\ell=1$, (\ref{4eq:MKdVh})
 turns out to be the MKdV equation (\ref{4eq:MKdV}).
These parameters $t_\ell$ correspond to
the Schwinger proper time in  field theory \cite[(3.5.11)]{Ramond};
it represents the freedom of the thermal fluctuation
which corresponds to quantum fluctuation in  
quantum field theory.

Let the subspace $\cM^{[\le g]}$ of $\cM$ be
$$
\cM^{[\le g]}:= \{Z\in \cM\ |\ \Omega^\ell \partial_s q = 0, \ell > g\}.
$$
We have a filtration in $\cM$, i.e., 
$\cM^{[\le g-1]}\subset \cM^{[\le g]}$.
Heuristically (for mathematical rigor we would need to identify the time flows
on the Jacobian of the spectral curves), we define 
$$
  \cM^{[g]}:= \cM^{[\le g]}\setminus \cM^{[\le g-1]}.
$$
As showed in \cite{MO03a}, $\cM^{[g]}$ 
corresponds to the moduli space of
 hyperelliptic curves of
genus $g$. Thus the density of states is decomposed into 
$$
\mathrm{Vol}(\cM_E) = \sum_{g=0}^\infty \mathrm{Vol}(\cM_E^{(g)}),
$$
where $\cM_E^{(g)}:=\cM_E\cap \cM^{(g)}$.

As is showed in subsection \ref{4sec:G01}, 
$\cM^{[0]}$ and $\cM^{[1]}$
are  singletons, i.e., sets with one element,
which are realized as the minimal points of the EB energy;
$\cM^{[0]}$ consists of a circle with  EB energy $E_{EB}^{(0)}=2\pi^2/L$,
 number of self-contact points $i^{(0)}=0$ and 
 winding number $w^{(0)}=1$, 
whereas the element of $\cM^{[1]}$ is the figure-eight
with $E_{EB}^{(1)}=(56.21980489\cdots)/L$,
number of self-contact points $i^{(1)}=1$ and 
 winding number $w^{(1)}=0$ \cite{WadatiTsuru}.
They could be the ground states, depending on the boundary conditions
given by  the winding number and the number of self-contact points
but the objective in the statistical mechanics of elastica 
is to evaluate the appearance probability of these shapes by
considering the energy (\ref{eq:cEZ}) and the temperature $1/\beta$.
When $\alpha_1=\alpha_2=0$,
$\mathrm{Vol}(\cM_E^{(g)})=\delta_{g,a}\delta(E-E_{EB}^{(a)})$ in
the neighborhood of $E_{EB}^{(a)}$ ($a=0,1$) where $\delta(x)$ is the
Dirac delta function.
For $E\le E_{EB}^{(1)}$, $\mathrm{Vol}(\cM_E^{(g)})$ 
is equal to zero for $g>1$ since 
the EB energy of an element in $\cM_E^{(g)}$ $g>1$ 
is much greater than 
$E_{EB}^{(1)}$:
the elements in $\cM_E^{(g)}$ $g>1$ cannot have  minimal energy
and must be excited states.
As mentioned below, toward the end of this section, 
our model naturally contains the Euler's elastica as $\beta \to \infty$.

The main theoretical issue in this paper is
 to find the elements in $\cM_E^{[g]}$ for $g>1.$
\bigskip

Given that the MKdV hierarchy is completely integrable \cite{Mat97, MO03a}, 
if we find a solution of (\ref{4eq:MKdVh}) in $\cM^{[g]}$ 
whose EB energy is $E$,
the time parameters $(t_1, t_2, \ldots, t_g) \in \RR^g$ must have periodicity
and have  fundamental domain $\Gamma_{g,E}$ 
producing a $g$-dimensional torus orbit.
In the orbit in $\Gamma_{g,E}$, the time deformations give
 the same EB-energy and thus
the volume of $\Gamma_{g,E}$ contributes to the density  states
$\mathrm{Vol}(\cM_E^{(g)})$.
Moreover, 
the solution uniquely depends on certain parameters $a:=
(a_1, a_2, \ldots, a_{m_g})$ in an $m_g$-dimensional subset 
$U \subset \RR^{m_g}$, 
which is related to the moduli space of the
hyperelliptic curves of genus $g$. 
The space $\Gamma_{g,E}$ depends on $a$, so we denote it
$\Gamma_{g,E}(a)$.
In general, different points in $U$ give  different
EB-energy but there exists a subset $U_E \subset U$ such that
the solution of the MKdV hierarchy has the energy $E$.
Then $\cM_E^{(g)}$  consists of these $U_E$ and $\Gamma_{g,E}(a)$
with a fiber structure $\pi: \cM_E^{(g)} \to U_E$ and 
$\pi^{-1}(a)=\Gamma_{g,E}(a)$ \cite{Mat97}.
In terms of these data.
 we may compute the partition functions $\cZ[\beta]$ and find the
distributions of shapes of elastica depending on the temperature $1/\beta$.

Here we give a comment on the case of  non-zero $\alpha_1$ and $\alpha_2$.
Equation (\ref{4eq:MKdVh}) is obtained by using certain local data in 
 function space, and the geometry of curvature,
and it is expected that every element $Z$ in $\Gamma_{g,E}(a)$ has the
same $i(Z)$ and $w(Z)$. Thus even for non-zero $\alpha_1$ and $\alpha_2$,
it is expected that 
the isometric and iso-energy condition provides the same 
equation (\ref{4eq:MKdVh});
this is obviously correct for $g=0$ and $g=1$.
However, it is not known even for $g=2,$ the case of this paper.

\bigskip
As a starting point, we
show the exact treatment of the statistical mechanics of elastica
for genus one and two, following \cite{Mat08}.
The shapes of the trajectories for
$g=0$ and $1$ are treated in subsection \ref{4sec:G01}
below.
Implicitly, we avoid the winding states of elastica.
However, $\cM$ could contain  winding elastica, cf.
\cite{Mat08}, e.g., $n$-times winding circles with radius $L/2\pi n$.
$\cM^{[0]}$ and $\cM^{[1]}$ have infinitely many elements with energy
$n^2 E_{EB}^{(0)}$ and $n^2 E_{EB}^{(1)}$. As a toy model,
the partition functions of these states is given by 
$$
\cZ^{(0,1)}[\beta] = \sum_{g=0}^1\sum_{n=1}^\infty 
\ee^{-\beta (n^2 E_{EB}^{(g)} + \alpha_2 n w^{(g)}) }
$$
and are expressed by the elliptic theta functions \cite{Mat08}.
(In the winding model,  the number of self-contact points 
 cannot be computed precisely 
for $n>1$ and thus we omitted it.)
Due to the difference between $E_{EB}^{(0)}$ and $E_{EB}^{(1)}$,
for given $\beta$, 
the appearance probability of these different shapes is obtained.
The comparison between
$n^2 E_{EB}^{(0)}$ and $n^2 E_{EB}^{(1)}$ is also studied in \cite{Sa},
in which the relation $E_{EB}^{(0)}< E_{EB}^{(1)}<4 E_{EB}^{(0)}$
is evaluated. In \cite{Sa}, the complete integrability of Euler's problem is
proven by control theory. It would be interesting to extend this approach to
higher genus.
When $\beta \to \infty$, it is obvious that the circle has minimal
energy if $\alpha_1=\alpha_2=0$. However since
the winding number $w^{(a)}$ is a topological invariant, 
 when $\alpha_2>0$ is sufficiently large 
(or $\alpha_1<0$ is sufficiently small if it is 
added only for $n=1$ to the energy), 
the figure-eight appears even for small temperature $1/\beta$ \cite{WadatiTsuru}.
Using  properties of  elliptic theta functions with the 
Poisson summation formula, we can find the analytic properties of 
$\cZ^{(0,1)}[\beta]$ described in \cite{Mat08}.

\bigskip

\section{Hyperelliptic solutions of generalized elastica}
\label{sec:HESGE}

We review the solution of the generalized elastica problem
 \cite{Mat02b,Mat07} for a hyperelliptic curve $X_g$
of genus $g$,
\begin{equation}
\left\{(x,y) \in \CC^2 \ |
\ y^2 = (x-b_1)(x-b_2)\cdots(x-b_{2g+1})\right\}
\cup \{\infty\},
\label{4eq:hypC}
\end{equation}
where $b_i \in \CC$ are mutually distinct complex numbers.
Let $\lambda_{2g}=\displaystyle{-\sum_{i=1}^{2g+1} b_i}$
and 
$S^k X_g$ be the $k$-th symmetric product of the curve 
$X_g$. The Abel integral
$v : S^k X_g \to \CC^g$, $(k=1, \ldots, g)$
is defined by its $i$-th component
$v_i$ $(i =1, \ldots, g)$,
$$
v_i((x_1,y_1),\ldots,(x_k,y_k))=\sum_{j=1}^l
 v_i(x_j,y_j), 
$$
\begin{equation}
v_i(x,y) = \int^{(x, y)}_\infty d u_i,\quad
d u_i = \frac{x^{i-1}d x}{2y}.
\label{4eq:firstdiff}
\end{equation}

\begin{theorem} {\textrm{\cite{Mat02b,Mat07}}}
\label{4th:MKdVloop}
For 
$((x_1,y_1),\cdots,(x_g,y_g))
\in S^g X_g$,
 a fixed branch point $b_a$
$(a=1, 2, \ldots, 2g+1)$,
and
$u:= v( (x_1,y_1),$ $\cdots,(x_g,y_g) )$,
$$
\displaystyle{
   \psi(u) :=
-\ii \log (b_a-x_1)(b_a-x_2)\cdots(b_a-x_g)
}
$$
satisfies the MKdV equation over $\CC$,
\begin{equation}
	(\partial_{u_{g-1}}-\frac{1}{2}
(\lambda_{2g}+3b_a)
          \partial_{u_{g}})\psi
           -\frac{1}{8}
\left(\partial_{u_g} \psi\right)^3
 -\frac{1}{4}\partial_{u_g}^3 \psi=0,
\label{4eq:loopMKdV2}
\end{equation}
where $\partial_{u_i}:= \partial/\partial u_i$
as an differential identity in $S^g X_g$
and $\CC^g$.
\end{theorem}

It is worthwhile noting 
 the difference between the MKdV equations
(\ref{4eq:MKdV}) over $\RR$ 
and (\ref{4eq:loopMKdV2})
over $\CC$.
By introducing real and imaginary parts,
$ u_b = u_{b\,\rr} + \ii u_{b\,\ri}$ and 
$ \psi = \psi_{\rr} + \ii \psi_{\ri}$,
the real part of (\ref{4eq:loopMKdV2}) is reduced to
the gauged MKdV equation
with gauge field 
$A(u)=(\lambda_{2g}+3b_a-
\frac{3}{4}(\partial_{u_{g}\, \rr}\psi_\ri)^2)/2$,
\begin{equation}
-(\partial_{u_{g-1}\, \rr}-
A(u)\partial_{u_{g}\, \rr})\psi_\rr
           +\frac{1}{8}
\left(\partial_{u_g\, \rr} \psi_\rr\right)^3
+\frac{1}{4}\partial_{u_g\, \rr}^3 \psi_\rr=0
\label{4eq:gaugedMKdV2}
\end{equation}
by the Cauchy-Riemann relations. 

\bigskip
In order to obtain a solution of
(\ref{4eq:MKdV}) or a generalized elastica
in terms of the data
in Theorem \ref{4th:MKdVloop}, 
the following conditions must be satisfied:

\begin{enumerate}

\item[CI] $|x_i - b_a|=$ a constant $: =\gamma> 0$ for all $i$ in 
Theorem \ref{4th:MKdVloop} \cite{Mat07},

\item[CII] $d u_{g\,\ri}=d u_{g-1\, \ri}=0$
in Theorem \ref{4th:MKdVloop} \cite{Mat07},

\item[CIII] $A(u)$ is a real constant:
if $A(u)=\gamma \varepsilon$ constant,
(\ref{4eq:gaugedMKdV2}) is reduced to (\ref{4eq:MKdV}). 
\end{enumerate}


\subsection{Genera zero and one }
\label{4sec:G01}

\begin{figure}
\begin{center}
\includegraphics[height=0.6\hsize, angle=90]{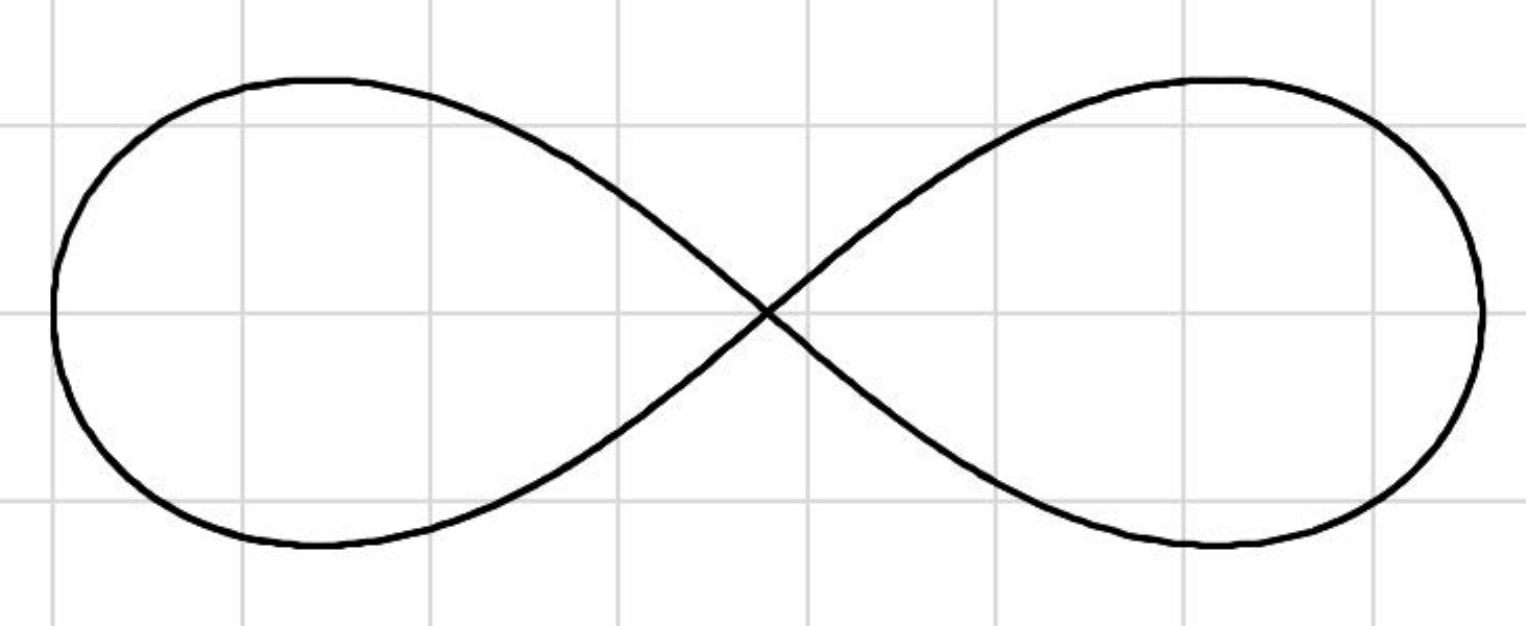}
\caption{The figure-eight}
\label{4fg:lsoliton_g1}.
\end{center}
\end{figure}

First we will consider genus one cases, including
genus zero,  studied by Euler and Bernoulli.
Using the isometric deformation $Z(\delta t)=Z(0) + \delta t \partial_t Z(0)$ 
($\partial_t Z(0)=(v^{(r)}+\ii v^{(i)})\partial_s Z$), after
the Goldstein-Petrich method \cite{GoldsteinPetrich1},
we consider the minimal problem for
the EB energy \cite{Mat10}. 
For the deformation,
we require
$\displaystyle{
\frac{\cE_{\rm{EB}}[Z(\delta t)]-\cE_{\rm{EB}}[Z(0)]}{v^{(r)} \delta t} = 0
}$ modulo $\delta t$. This yields  the static MKdV equation,
$
a \kappa +\frac12 \kappa^3+\partial_s^2 \kappa =0,
$
associated with the elliptic curve
$
(\partial_s \kappa)^2 + \frac{1}{4} \kappa^4 + a \kappa^2 + b = 0,
$
where $a$ and $b$ are integral constants \cite{Mat10}.
Let us consider the elliptic curve $X_1$ 
($\hat y=2y$ in (\ref{4eq:hypC}) $g=1$)
in the Weierstrass standard 
form \cite{WhittakerWatson},
\begin{equation}
\hat{y}^2=4(x-b_1)(x-b_2)(x-b_3),
\label{4eq:curve-g1}
\end{equation}
where
$b_1 = -\frac{1}{6}a$,
$b_2 =  \frac{1}{12}a+\frac{1}{4}\sqrt{b}$, and
$b_3 =  \frac{1}{12}a-\frac{1}{4}\sqrt{b}$
satisfying  $a^2-b=16$.
We identify the coordinate $u=(u_1)$ of the complex plane $\CC$ 
with $u_1$ and $u=v_1(x,y)$ for $(x,y) \in X_1$.

\begin{lemma}
Let $\ee^{2\ii\varphi} :=(x-b_1)/\gamma$ for 
$e_{c-1} := b_c - b_1$ and $\gamma=\sqrt{e_{1}e_{2}}$.
The  holomorphic one-form (\ref{4eq:firstdiff}) is
\begin{equation}
d u = \frac{ 2k\,  d\varphi}
	 {
           \sqrt{1-k^2 \sin^2 \varphi}},
      \label{eq:am-g1}
\end{equation}
where $\gamma = 1$, $(\sqrt{e_{2}}-\sqrt{e_{1}})^2=(a-4)/2$
and  the modulus
$\displaystyle{
k :=\frac{2\ii \root4\of{e_{2}e_{1}}}
{\sqrt{e_{2}}-\sqrt{e_{1}}}=
\sqrt{\frac{8}{4-a}}
}$.
\end{lemma}

In the cases of genera zero and one,
it is easy to impose the conditions CI-III, i.e.,
$\partial_{u \rr}\psi_\ri=0$, $d u_{\ri}=0$ by $|x-b_1|$ $=\gamma$, and 
therefore we are concerned with $\phi = \psi_\rr=2\varphi_\rr$,
which obeys (\ref{4eq:MKdV}) and (\ref{4eq:gaugedMKdV2}).

The solutions are expressed by the Jacobi elliptic functions, cf.,  
e.g. \cite{Love, Sa, WadatiTsuru}; 
therefore, they can instead be written in terms of
 the Weierstrass elliptic functions \cite{Mat10};
the latter can be generalized to higher genus 
 and they are briefly reviewed 
in the introduction to Section \ref{sec:HESGE} \cite{Mat02b,Mat10}.
In this paper, in order to evaluate these solutions 
numerically,
we show that Euler's figure eight is
directly expressed by means of   Euler's method,
though current mathematical software easily 
reproduces Euler's original results;
our computation in Figure \ref{4fg:lsoliton_g1}
shows a simple numerical method suffices.


\begin{enumerate}
\item{$g=0$ case:}\ 
We consider the limit of 
$k\to 0$ with $2k d \varphi=d s$. By identifying
$d u$ with $d s$, we have the circle as the minimal path of
the EB energy.

\item{$g=1$ case:}\ 
We let $d u = d s$, $k=2.398107502$
for $a=2.608918126$.
We numerically evaluate the figure-eight curve
of in Figure~\ref{4fg:lsoliton_g1},
as illustrated in Euler's book 
\cite{Euler44}: here we plot 
$Z(s)=(X + \ii Y)(s)$ 
as a function of $s$.
We 
used Euler's approximation method,
$ \varphi(s + \delta) 
\approx \varphi(s)+
\frac{\partial \varphi}{\partial s}\delta$
and
$Z(s+\delta) =
Z(s)+\ee^{2\ii \varphi} \delta
$ for a small number $\delta$.
In the computations, $\varphi$ moves back and forth
in the interval $[\varphi_s, \varphi_e]$
where $\varphi_s = \sin^{-1}(k^{-2})\in [0, \pi/2)$ and 
$\varphi_e = \pi - \varphi_s$.

\end{enumerate}

\subsection{Genus Two}\label{sec:g=2}

In this subsection, we investigate the conditions CI-III
for hyperelliptic curves $X_2$ of genus $g = 2$.
It turns out that in order to obtain the solution of 
(\ref{4eq:MKdV}) based on Theorem \ref{4th:MKdVloop}, 
we need higher-genus hyperelliptic curves ($g>2$). 
However, we show that even data in  $S^2 X_2$ give
 trajectories for 
generalization of elastica cf.  Figure
\ref{4fg:lsoliton_g2} and Figure \ref{4fg:lsoliton_g22}.


We choose coordinates $u = {}^t(u_1, u_2)$
 in  $\CC^2$;
$u_i = u_i^{(1)}+u_i^{(2)}$ where $u_i^{(j)}
=v_i((x_j, y_j))$.

We let $a=5$ in Theorem \ref{4th:MKdVloop}, $b_5=-\gamma=-1$,
and $e_{c} := b_c - b_5$ $(c=1,2,3,4)$.

We restrict
the moduli (or parameter) space of the curve $X_2$
by the following:


\noindent
{\bf{Conditions:}}
$\sqrt{e_{2a-1}} = \alpha_a +\ii \beta_a$,
$\sqrt{e_{2a}} = \alpha_a -\ii \beta_a$
where $\alpha_a, \beta_a\in \RR$, $a,b =1,2$, satisfying 
$\alpha_a^2 + \beta_a^2 = \gamma$.


Under this assumption, we have the natural
extension of genus-one elastica; indeed, direct computation shows the 
following:

\begin{lemma} \label{4lm:g2gene_y2}
Let $\gamma \ee^{2\ii\varphi} :=(x-b_5)$,
(\ref{4eq:hypC}) $g=2$ is
\begin{equation}
y^2=16 \frac{\gamma^5\ee^{6\ii\varphi}}{k_1^2 k_2^2} 
(1-k_1^2 \sin^2 \varphi)(1-k_2^2 \sin^2 \varphi)
\end{equation}
where 
$\displaystyle{
k_a = \frac{2\ii\sqrt[4]{e_{2a-1}e_{2a}}}{\sqrt{e_{2a-1}}- \sqrt{e_{2a}}}
=\frac{\sqrt{\gamma}}{\beta_a}}$, $(a=1,2)$.
\end{lemma}

We consider a point $((x_1, y_1),(x_2,y_2))$ in $S^2 X_2$
under the condition CI, $|x_c-b_5|=\gamma$.
We define the variable $\varphi_a$ by $x_c= \gamma \ee^{\ii \varphi_c}
(\ee^{\ii \varphi_c}+ (b_5/\gamma) \ee^{-\ii \varphi_c})$
$(c=1,2)$.
Noting
$d x_c = 2 \ii \gamma \ee^{2\ii \varphi_c}d\varphi_c$ and
 $x_c d x_c = -4 \gamma \ee^{3\ii \varphi_c}\sin \varphi_c\ d \varphi_c$,
we have the holomorphic one forms
$(d u_1^{(c)}, d u_2^{(c)})$ $(c=1,2)$, 
\begin{equation}
\left(
\frac{ 
(\sin \varphi_c+ \ii \cos\varphi_c)\ d \varphi_c}{
2 \gamma K(\varphi_c)},
-
\frac{ \sin \varphi_c\ d \varphi_c}{
K(\varphi_c)}\right),
\end{equation}
where $\displaystyle{
K(\varphi):=\frac{\sqrt{\gamma
 (1-k_1^2 \sin^2 \varphi)(1-k_1^2 \sin^2 \varphi)}}
{k_1k_2}}$.

\begin{lemma} \label{4lm:dudphi}
Let $K_c:=K(\varphi_c)$. The following holds:

\begin{enumerate}

\item 
$
\displaystyle{
\left(\begin{array}{c}
d u_1 \\ d u_2
\end{array}\right)
=
\left(
\begin{array}{cc}
\frac{\ii \exp(-\ii\varphi_1)}{2\gamma K_1}&
\frac{\ii \exp(-\ii\varphi_2)}{2\gamma K_2}\\
\frac{- \sin(\varphi_1)}{K_1}&
\frac{- \sin(\varphi_2)}{K_2}
\end{array}\right)
\left(\begin{array}{c}
 d \varphi_1 \\ d \varphi_2
\end{array}\right)
}
$,

\item
$\displaystyle{
\left(\begin{array}{c}
 \partial_{ u_1} \\ \partial_{ u_2}
\end{array}\right)
=
\frac{2\gamma K_1 K_2}{\ii \sin(\varphi_2- \varphi_1)}
\left(\begin{array}{cc}
-\frac{\sin(\varphi_2)}{K_2}&
\frac{ \sin(\varphi_1)}{K_1}\\
-\frac{\ii \exp(-\ii\varphi_2)}{2\gamma K_2}&
\frac{\ii \exp(-\ii\varphi_1)}{2\gamma K_1}
\end{array}\right)
\left(\begin{array}{c} 
\partial_{ \varphi_1 } \\ \partial_{ \varphi_2 }
\end{array}\right)
}$.

\end{enumerate}

\end{lemma}

Due to Lemma \ref{4lm:dudphi} 
and the expression
$ \varphi_c = \varphi_{c\rr} + \ii \varphi_{c\ri}$, we have
\begin{equation}
\frac{\partial \varphi_{1\ri}}
{\partial u_{2\, \rr}}
=
\frac{K_1\sin \varphi_{2\rr}}
{\sin(\varphi_{1\rr}-\varphi_{2\rr})},\quad
\frac{\partial \varphi_{2\ri}}
{\partial u_{2\, \rr}}=
\frac{K_2\sin \varphi_{1\rr}}
{\sin(\varphi_{2\rr}-\varphi_{1\rr})}
\label{4eq:psii}
\end{equation}
and $\psi = 2 (\varphi_1 + \varphi_2)$ 
satisfies (\ref{4eq:loopMKdV2})
and (\ref{4eq:gaugedMKdV2}).

\bigskip



\begin{figure}
\begin{center}

\includegraphics[width=0.7\hsize]{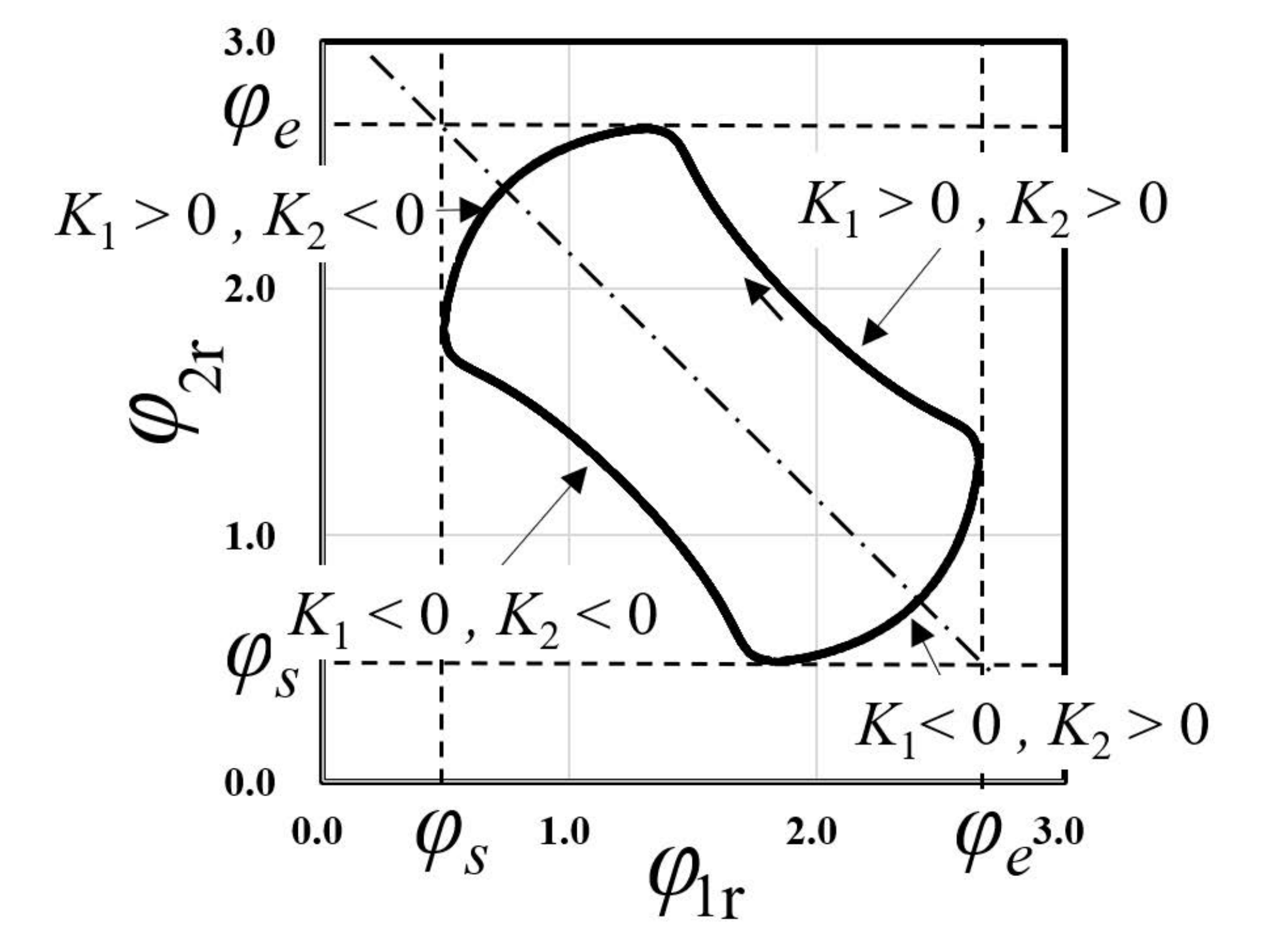}

\caption{The orbit $C$ of $\partial_{u_{2}\, \rr}\psi_\ri=$constant:
$\varphi_{1\rr}$ depends on $\varphi_{2\rr}$ for
Figure\ref{4fg:lsoliton_g2} (c). The arrow shows the
direction of $d\eta$.
}
\label{4fg:phi_phi}
\end{center}
\end{figure}

\begin{figure}
\begin{center}

\includegraphics[width=0.8\hsize]{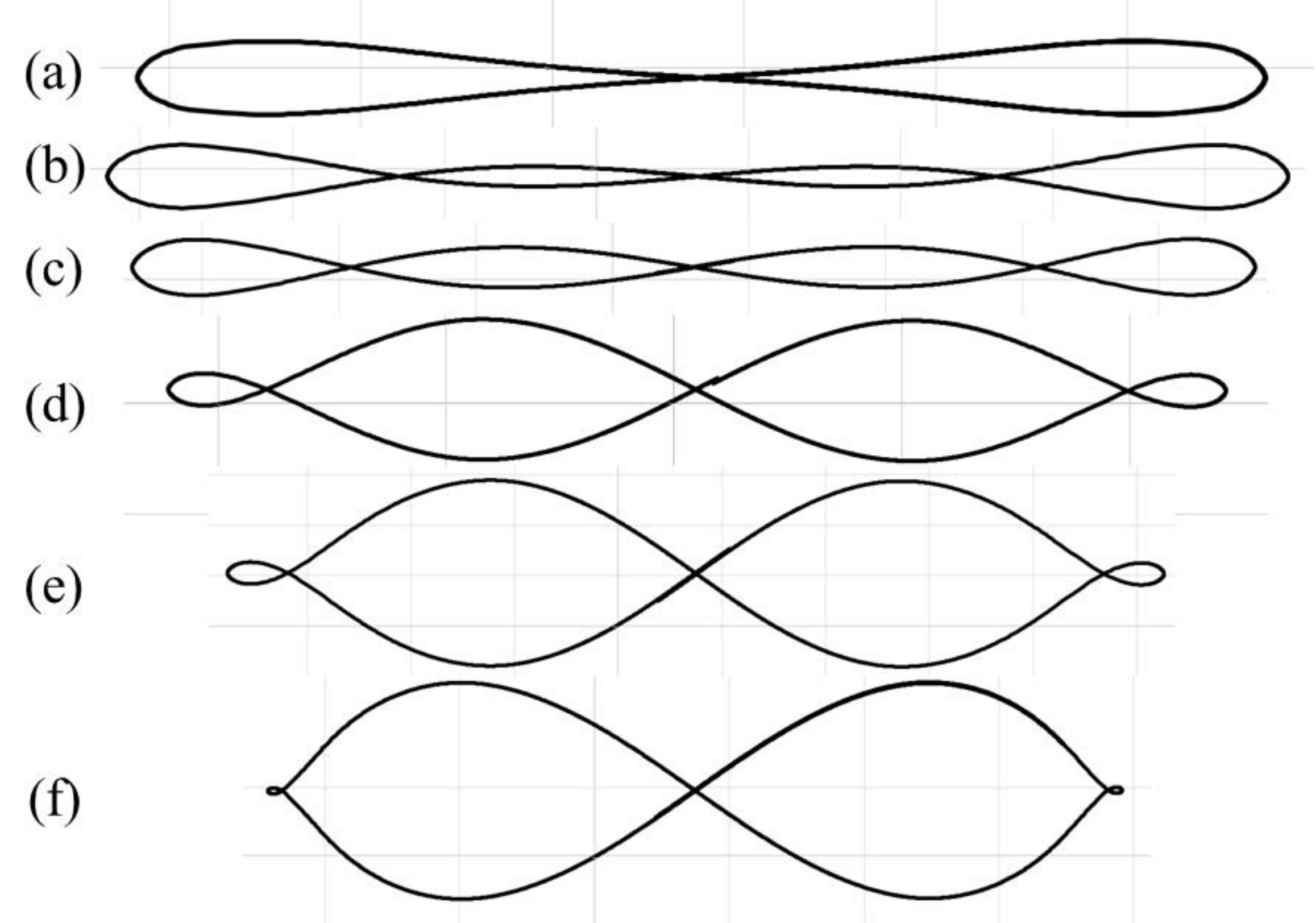}

\caption{Generalized elastica of genus two, where 
$(k_1, k_2)$ are as follows:
(a):(1.60,2.20), (b):(1.90,2.15),
(c) $(2.08, 2.32)$, (d): $(3.08, 3.32)$, (e): $(3.80. 4.20)$,
(f):(4.00,10.00).
}
\label{4fg:lsoliton_g2}
\end{center}
\end{figure}


\begin{figure}
\begin{center}
\includegraphics[width=0.8\hsize]{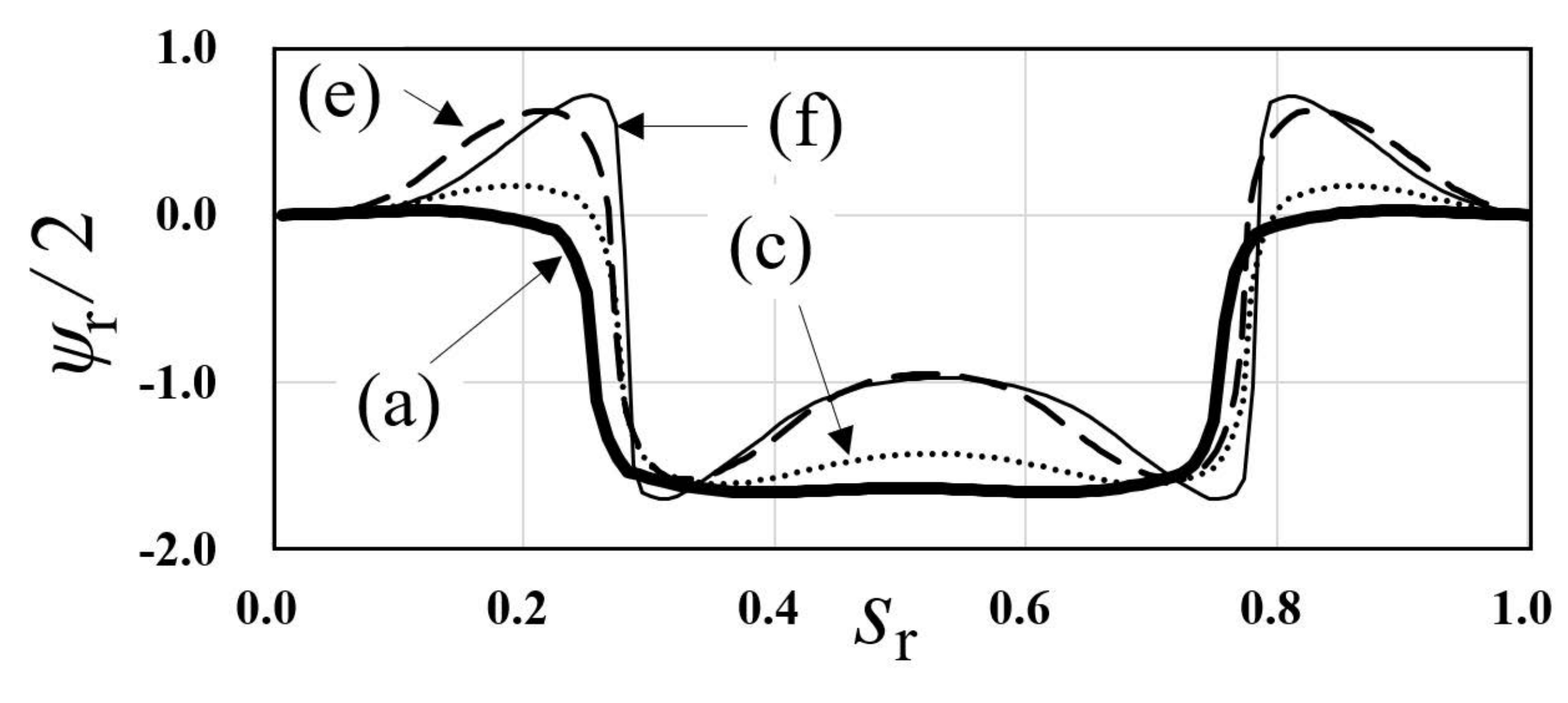}

\caption{$\psi_\rr/2$ of Figure~\ref{4fg:lsoliton_g2} (a), (c), (e), and (f).
}
\label{4fg:phi}
\end{center}
\end{figure}

\begin{figure}
\begin{center}
\includegraphics[width=0.8\hsize]{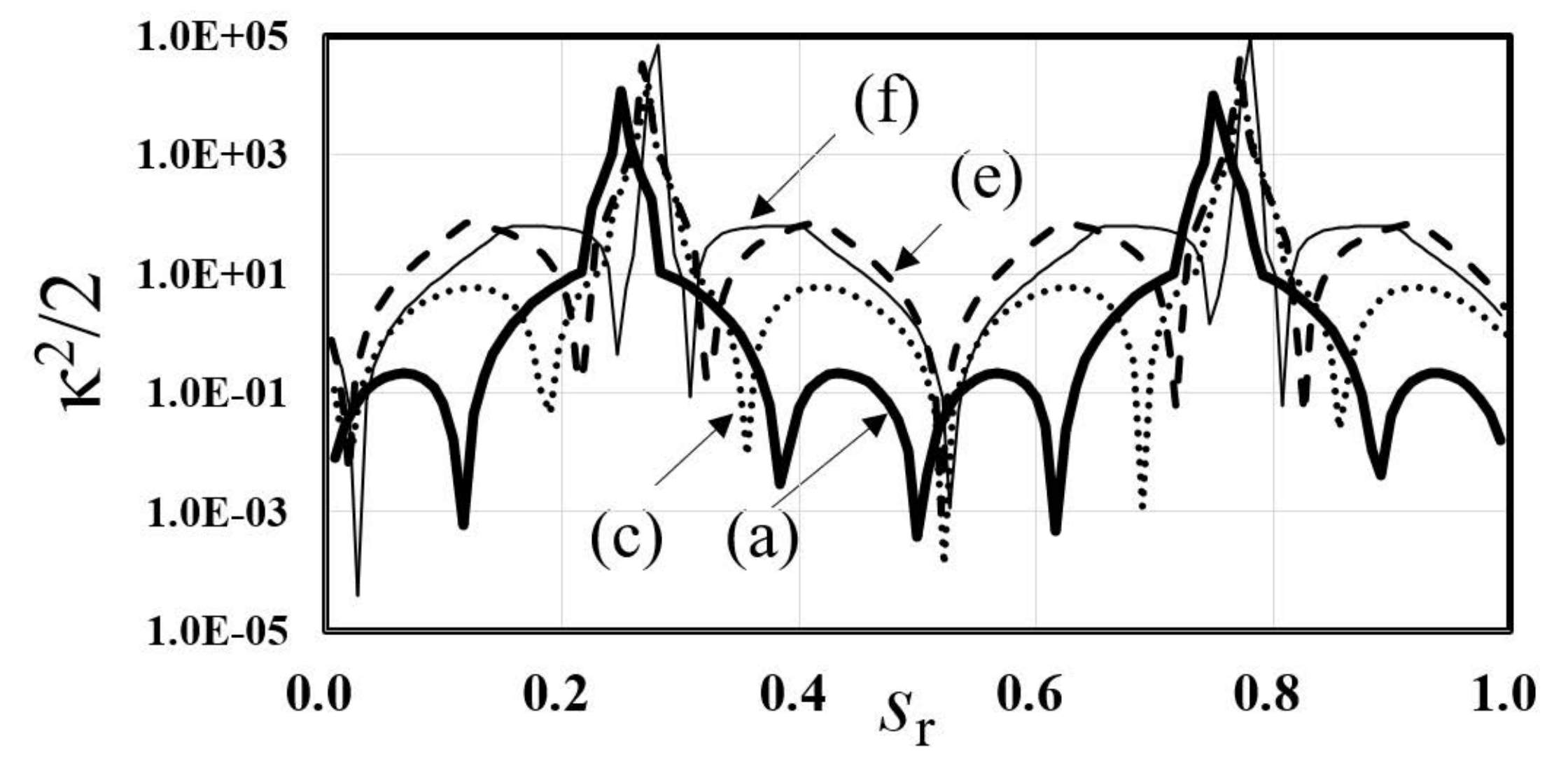}

\caption{$\kappa^2/2$ of Figure~\ref{4fg:lsoliton_g2} (a), (c), (e), and (f).
}
\label{4fg:kappa2}
\end{center}
\end{figure}

\begin{figure}
\begin{center}
\includegraphics[width=0.8\hsize]{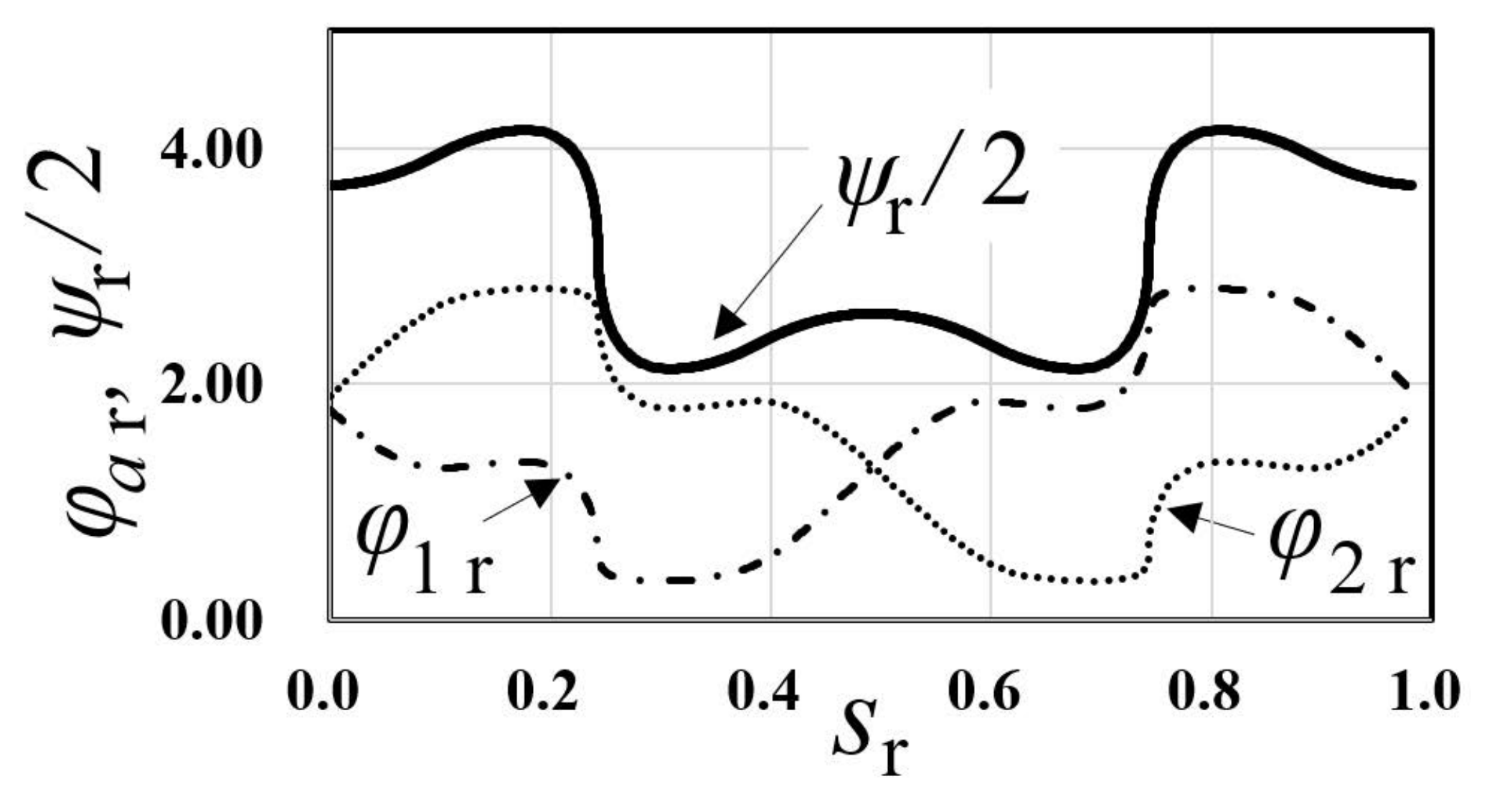}

\caption{The decomposition of $\psi_\rr/2$ to 
$\varphi_{1\rr}$ and $\varphi_{2\rr}$ of Figure~\ref{4fg:lsoliton_g2} (d).
}
\label{4fg:phi_psi}
\end{center}
\end{figure}


Let us investigate the conditions CII and CIII.
Assume $A(u)=\varepsilon\gamma$; then,
${
\left(\begin{array}{c}
\partial_t \\ \partial_s 
\end{array}\right)
=
\left(\begin{array}{cc}
 -4 & -4\varepsilon\gamma \\ 0 & 1 
\end{array}\right)
\left(\begin{array}{c}
\partial_{u_1} \\ \partial_{u_2} 
\end{array}\right)
}$
in (\ref{4eq:gaugedMKdV2})
shows the correspondence between the Abelian parameters 
$(u_1, u_2)$ and $(t,s)$ in the MKdV equation
(\ref{4eq:MKdV})
via 
${
\left(\begin{array}{c} 
d t \\ d s 
\end{array}\right)
=
\left(
\begin{array}{cc} -1/4 & 0 \\ \varepsilon\gamma & 1 \end{array}
\right)
\left(\begin{array}{c}
d u_1 \\ d u_2
\end{array}\right)
}$.
Using this expression,
$d t = d t_\rr + \ii d t_\ri$ and
$d s = d s_\rr+ \ii d s_\ri$,
we have the following lemma:
\begin{lemma}
$\displaystyle{
\left(\begin{array}{c}
d t_\rr \\ d t_\ri \\ d s_\rr\\ d s_\ri
\end{array}\right)
= M
\left(
\begin{array}{cc}
d\varphi_{1\rr} \\ d\varphi_{1\ri} \\ 
d\varphi_{2\rr} \\ d\varphi_{2\ri} 
\end{array}\right)}$
where
%
\begin{equation}
M:=
\left(
\begin{array}{cccc}
-\frac{\sin\varphi_1}{8\gamma K_1} &
\frac{\cos\varphi_1}{8\gamma K_1} &
-\frac{\sin\varphi_2}{8\gamma K_2} &
\frac{\cos\varphi_2}{8\gamma K_2} \\
-\frac{\cos\varphi_1}{8\gamma K_1} &
-\frac{\sin\varphi_1}{8\gamma K_1} &
-\frac{\cos\varphi_2}{8\gamma K_2} &
-\frac{\sin\varphi_2}{8\gamma K_2} \\
\frac{(2+\varepsilon)\sin\varphi_1}{2 K_1} &
-\frac{\varepsilon\cos\varphi_1}{2 K_1} &
\frac{(2+\varepsilon)\sin\varphi_2}{2 K_2} &
-\frac{\varepsilon\cos\varphi_2}{2 K_2} \\
\frac{\varepsilon\cos\varphi_1}{2 K_1} &
\frac{(2+\varepsilon)\sin\varphi_1}{2 K_1}&
\frac{\varepsilon\cos\varphi_2}{2 K_2} &
\frac{(2+\varepsilon)\sin\varphi_2}{2 K_2} \\
\end{array}\right).
\label{4eq:Mlsol}
\end{equation}
\end{lemma}

Due to condition CI, $d\varphi_{1\ri}=d\varphi_{2\ri}=0$, 
and thus we have $d t_\rr\propto d s_\rr$.

We consider the condition CIII, i.e., 
an orbit $C$ of constant   
$\partial_{u_2\rr} \psi_\ri$ 
in the $\varphi_{1\rr}$-$\varphi_{2\rr}$ plane
by requiring
\begin{equation}
d\left(\frac{\partial \psi_{\ri}}
{\partial u_{2\, \rr}}\right)=
\frac{\partial (\partial_{u_2\rr}\psi_{\ri})}
{\partial \varphi_{1\rr}}d\varphi_{1\rr}
+
\frac{\partial (\partial_{u_2\rr}\psi_{\ri})}
{\partial \varphi_{2\rr}}d\varphi_{2\rr}=0,
\end{equation}
using (\ref{4eq:psii}).
By parameterizing $C$ by $\eta\in \RR$, we have
the curve equation,
\begin{equation}
d\varphi_{1\rr}=
\frac{\partial (\partial_{u_2\rr}\psi_{\ri})}
{\partial \varphi_{2\rr}} d \eta, \quad
d\varphi_{2\rr}=
-\frac{\partial (\partial_{u_2\rr}\psi_{\ri})}
{\partial \varphi_{1\rr}} d \eta, 
\label{eq:dvarphi_a}
\end{equation}
On $C$, $\psi$, $s$ and $t$ are functions of 
$\eta\in \RR$.
This implies that $d s_{\ri} \neq 0$ in general.

Due to
the number  of conditions CI, CII, and CIII
and the degrees of freedom of the real parameters
$u_1$ and $u_2$,
the consistency between CII and CIII is crucial.

We obtain $C$ by numerically solving 
(\ref{eq:dvarphi_a}) as in Figure \ref{4fg:phi_phi}
using Euler's method.
In the computations, $\varphi_a$ moves back and forth
in the interval $[\varphi_s, \varphi_e]$
where $\varphi_s = \sin^{-1}(k_1^{-2})\in [0, \pi/2)$ and 
$\varphi_e = \pi - \varphi_s$.
Since $\varphi_s$ and $\varphi_e$ are branch points in $X_2$
(the interval $[\varphi_s, \varphi_e]$
corresponds to one of the homology basis of $X_2$),
the sign of $K_a$ changes at the turns.

Due to multiplication by $M$ in (\ref{4eq:Mlsol}), $d s_{\rr}$ vanishes 
on the line
$L: \varphi_{1\rr}=\pi-\varphi_{2\rr}$ 
in the $\varphi_{1\rr}$-$\varphi_{2\rr}$ plane.
For our $g=2$ case, there exist
 crossing points between the line
$L$ and the orbit $C$
(Figure \ref{4fg:phi_phi}), and 
at the points $d s_{\rr}$ vanishes whereas $d s_{\ri}$ (or 
$d u_{2\ri}$) does not. Thus 
$\displaystyle{
\frac{d \psi_{\rr}}{d s_{\rr}} = \frac{\partial \psi_{\rr}}{\partial s_{\rr}}
-\frac{\partial \psi_{\rr}}{\partial s_{\ri}}\frac{d s_\ri}{d s_{\rr}}
}$
diverges at the points. 
Since $g=2$,
the crossing occurs for any parameters $k_c$ $(c=1,2)$, 
the divergence shows that
 the conditions CII and CIII are contradictory,
and thus we cannot find the solution of 
(\ref{4eq:MKdV}) based on Theorem \ref{4th:MKdVloop}
in genus two. To satisfy the conditions, we need higher-genus 
hyperelliptic curves $X_g$ $(g>2)$.

\bigskip

However, the pole behavior of $d \psi_{\rr}/d s_{\rr}$ is of type
$1/\sqrt{s_\rr-s_{\rr0}}$ at $s_\rr=s_{\rr0}$ up to a constant 
factor. Therefore $\psi_\rr$ can be 
defined as a hyperelliptic (2-branch) function of $s_{\rr}$ using the data in 
Theorem \ref{4th:MKdVloop} along $C$. 
Since for finite $\delta \eta$, we obtain
$\delta\varphi_{c\rr}$ $(c=1,2)$
by (\ref{eq:dvarphi_a}),
we compute $\delta s_\rr$ by using $M$ in (\ref{4eq:Mlsol}), 
and 
approximate $Z(s_\rr + \delta s_\rr)
\approx Z(s_\rr)+
\ee^{2(\varphi_{1\rr}+ \varphi_{2\rr})\ii}
\delta s_\rr$ using  Euler's method
for appropriate initial data $\varphi_a(s=0,t=0)$
so that $Z$ is periodic in $s_\rr$;
since $Z$ is in general not periodic,
we determine appropriate initial states  by  
the shooting  method with the bisection process.

Finally, we numerically obtain 
the shapes of a generalization of the elastica 
of genus two in 
Figure~\ref{4fg:lsoliton_g2}. Though they are not
solutions of (\ref{4eq:MKdV}),
they identically satisfy (\ref{4eq:loopMKdV2})
and are a natural extension of Euler's results.
We also numerically compute and display
their $\psi_\rr$ in Figure~\ref{4fg:phi}
and the energy density $\kappa^2/2$ in Figure~\ref{4fg:kappa2}.

\subsection{Comparison of elastica's periodic orbits and
  DNA}\label{comparison}
We offer a few remarks that provide evidence, for the shapes that we obtained,
to actually exhibit the geometry of supecoiled DNA.

\bigskip
Coleman and Swigon classify the  equilibrium configurations of 
a knot-free loop \cite[Figure 1]{CS}; Arnold gives a complete list
 using topological indices
\cite[Figure 11]{Arnold}. Figure~\ref{4fg:lsoliton_g2} (b)-(f) 
corresponds to (d) in the
Coleman-Swigon list \cite{CS}
and $(0,0,-3)$ in Arnold's list \cite{Arnold} whereas
Figure~\ref{4fg:lsoliton_g2} (a) and Figure~\ref{4fg:lsoliton_g1}
correspond to (b) in \cite{CS} and $(0,0,-1)$ in \cite{Arnold} respectively.
The last entries $-3$ and $-1$ in 
Arnold's symbol correspond to the number of 
self-contact points with minus sign, $-i(Z)$ in this case.
It is noted that
the shapes in Figure~\ref{4fg:lsoliton_g2} have the same
zero winding number
as the figure-eight, $w^{(1)}=0$.

Since the shapes in
Figure~\ref{4fg:lsoliton_g2} satisfy (\ref{4eq:loopMKdV2})
for genus two
(though  they are not
solutions of (\ref{4eq:MKdV})), we can bring them as evidence 
because they 
reproduce some  properties of the AFM or 
EMS images of the supercoiled DNA 
\cite{BVL, Betal,KB, LS, VV};
Figure 24-8 in \cite{VV} displays the five electron micrographs of DNA
(mini ColE1 plasmid dimer, 5kb) by Laurien Polder
 \cite[p.36, Figure 1-19]{KB}, 
which shows the supercoiling from fully relaxed to tightly coiled.
The two of them which have four and six self-contact points,
i.e., $(0, 0, -4)$ and $(0, 0, -6)$ in Anold's symbol, with
{\lq}weak elastic forces{\rq},
  resemble  Figure~\ref{4fg:lsoliton_g2} (b) and (c),
though the number of self-contact points differ.
As in the AFM images of mini plasmid, 688bp
 \cite[Figures 1, 2]{LS}, we find distorted circle and figure-eight,
i.e., $(0, 0, 0)$ and $(0, 0, -1)$ in Anold's symbol,
both correspond to the solutions in elliptic functions 
in subsection \ref{4sec:G01} and Figure~\ref{4fg:lsoliton_g2} (a).
On the other hand, the large plasmid 
in \cite[Figure 3]{Betal}, \cite[Figures 2 and 3]{BVL},
\cite[Figures 1 and 2]{LS},
and figure \cite{KB,VV}
have more complicated shapes with 
larger self-contact numbers.
Some of them are the loops with Anold's symbol $(0, 0, -\ell)$,
 $(\ell \ge 5)$ with 
{\lq}weak elastic forces{\rq},
and 
Figure~\ref{4fg:lsoliton_g2} (b)-(f)  reproduce such properties
though they pertain to the $\ell = 3$ case

The $\psi_\rr$ in Figure~\ref{4fg:phi} exhibits 1)  periodicity
and 2) mixing of  so-called kink (anti-kink) type 
mode and breather type mode \cite{AS}; due to the kink type mode,
the energy density (or elastic force) is localized (cf.  
Figure~\ref{4fg:kappa2})
and has a singularity; the local extrema correspond to
the two points of maximal curvature of the loops in 
Figure~\ref{4fg:lsoliton_g2}.

The mixing of both modes turns out to be
 a higher-genus phenomenon, not present in genus one.
The $\psi_\rr$ in Figure~\ref{4fg:phi} consists of
$\varphi_{1\rr}$ and $\varphi_{2\rr}$ and 
the addition of both form the mixing; we show it for
the case (c)  in Figure~\ref{4fg:phi_psi}.
Since genus 2 gives two parameters $\varphi_{a\rr}$, there are various (non
unique) types of mixing and we have 
the family of shapes in Figure~\ref{4fg:lsoliton_g2},
which reproduce the shapes of DNA.

\subsection{Considerations on the energy of the orbits}

To reproduce the shapes of the supercoiled DNA mathematically,
several models are proposed (cf. Introduction) based on Euler's elastica.
However, in these approaches even  three dimensional effects such as twist and 
other perturbative effects cannot exhibit the mixing.
Indeed, we consider the  minima of the energy
as in subsection~\ref{4sec:G01}:
by fixing the boundary conditions 
(the winding number and the number of the self-contacts),
 the minimum
 should be uniquely determined.
Since the image of $v$  in (\ref{4eq:firstdiff})
for $g=1$ is one dimensional,
the elliptic functions represent only one of the kink 
solution or  breather-type solution, and 
as in  subsection~\ref{4sec:G01},
 the shape is rather simple.

In \cite{CS, TMT}, the writhing modes were demonstrated using a real 
thick rubber rod in \cite[Figure 8]{TMT} and numerical elastic rods 
with finite thickness \cite[Figures 7 and 8]{CS} which must preserve 
the self-contacts. However
if one takes the limit of the rod's thickness to zero 
to model Euler's elastica, 
the loops must collapse. 
This contradicts the observation in \cite{BVL, LS, VV};
In general, the supercoiled DNAs 
have  shapes whose self-intersection does not change even taking
the limit of their thickness to zero.
Therefore, shapes with  minimal energy and topological constraints
do not recover the shapes of supercoiled DNAs in general,
and we have to consider additional effects such as the thermal effect.
This results in the MKdV equation (\ref{4eq:MKdV}) and MKdV hierarchy
(\ref{4eq:MKdVh}). The associated  Riemann surfaces of higher genus
demonstrate the structure of DNA, which requires excited states.

Our results in Figure \ref{4fg:lsoliton_g2} provide
such shapes with the number of self-contact numbers $i(Z)>1$
with
{\lq}weak elastic forces{\rq} like observed DNA. We computed them in genus
two. 
We expect that
the closed orbits of generalized elastica with  higher genus have a larger
number of  
contact points, and faithfully reproduce the observed DNA shapes.

As mentioned in Section \ref{sec:SME},
on a physical model of DNA,
the EB energy is a component of the total energy (\ref{eq:cEZ}).
Since the singularity of the energy density is of type $d
s_{\rr}/(s_{\rr}-s_{\rr0})$, 
 it can be removed by  introducing a cut-off parameter.
Thus by fixing the cut-off parameter,
we may propose the shapes in Figure~\ref{4fg:lsoliton_g2} as a physical 
model: for example, 
we may regard Figure~\ref{4fg:lsoliton_g2} as a transition of shapes
from (a) to (f) (and further to figure-eight) 
depending on the variation of the temperature and
the difference of their energy; like $\cZ^{(0,1)}[\beta]$ in the $n=1$ case
in Section \ref{sec:SME}, the number of self-contact points  
have effects on the transition depending on $\alpha_1$
in (\ref{eq:cEZ})
though $\alpha_2$, the coefficient of the winding number
 in (\ref{eq:cEZ}), does not have any effect on it.

If we extend our model to three dimensions 
\cite{C0, C, GoldsteinLanger, Mat99a},
our results in Figure~\ref{4fg:lsoliton_g2}
naturally bring in the writhing numbers,
which are part of  the geometry of DNA.
When we consider the statistical mechanics of
elastica in three-dimensional space,
their configurations are given by
the  torsion $\tau(s)$ and the curvature $\kappa(s)$
\cite{Mat99a}.  The complex curvature 
$\displaystyle{\kappa_c = \kappa \exp\left(\int^s \tau(s) d s\right)}$
naturally appears.
If we employ the EB energy using $\kappa^2 = \overline{\kappa_c}\kappa_c$, 
the NLS and CmKdV hierarchies
provide the isometric and the iso-energy deformations
instead of the MKdV hierarchy for the elastica in a plane.
Several elastic models in three dimensional space are proposed.
Since the topological invariant for the strip is given by the linking
number $\mathrm{Lk}$ which consists of the writhing number 
$\mathrm{wr}$ and the twisting number $\mathrm{tw}$, i.e.,
$\mathrm{Lk}=\mathrm{wr}+\mathrm{tw}$,
these topological properties 
directly relate to the shapes;
for example, if we consider them as global effects, or
define the total energy 
$E_{\mathrm{total}}=E_{\mathrm{EB}}+\alpha_3 \mathrm{wr}
+\alpha_4\mathrm{tw}$ for 
coupling constants $\alpha_3$ and $\alpha_4$, they influence the 
distributions of shapes
as mentioned in the toy model $\cZ^{(0,1)}[\beta]$ in Section
\ref{sec:SME}. 
The models depend on how we treat these effects of the writhing and
twisting locally by certain elastic forces (by coupling with the
 torsion $\tau(s)$ and the curvature $\kappa(s)$) and/or globally.
When we employ the simplest energy $E_{\mathrm{EB}}$
even in the three dimensional model for simplicity, 
the CmKdV and the NLS equations,
which are complex valued \cite{EEK, P0, Shin}, appear instead of 
the MKdV equation \cite{C, GoldsteinLanger, Mat99a}.
Due to the complex-valued setting, we expect
 that the difficulty in solving the equations identified in this paper 
might be avoided even for $g=2$
and the writhing-like shapes in Figure~\ref{4fg:lsoliton_g2}
might be realized as real writhing in three-dimensional space,
by exactly solving the PDE.

\bigskip

\begin{figure}
\begin{center}
\includegraphics[height=0.38\hsize, angle=0]{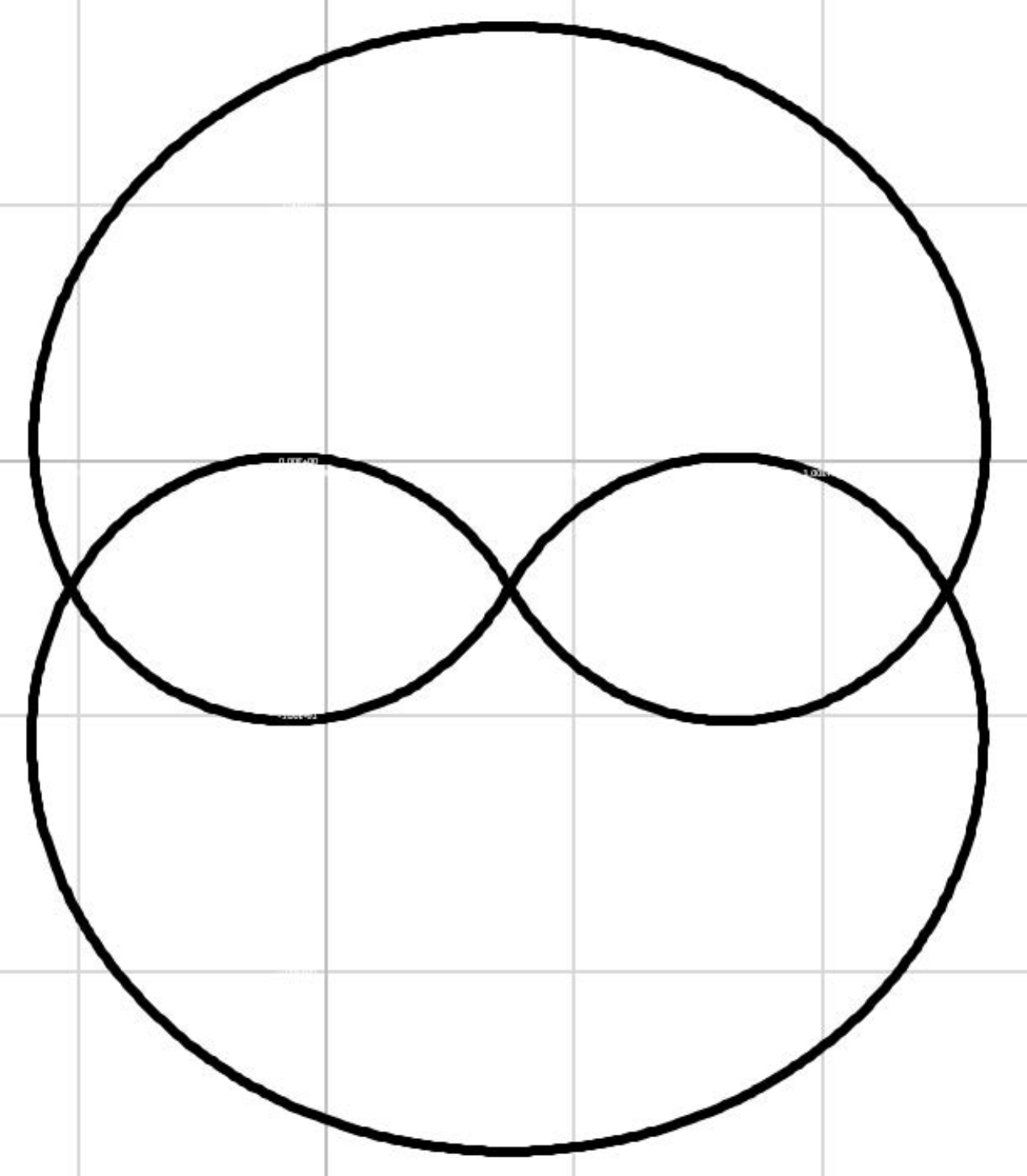}
\includegraphics[height=0.4\hsize, angle=0]{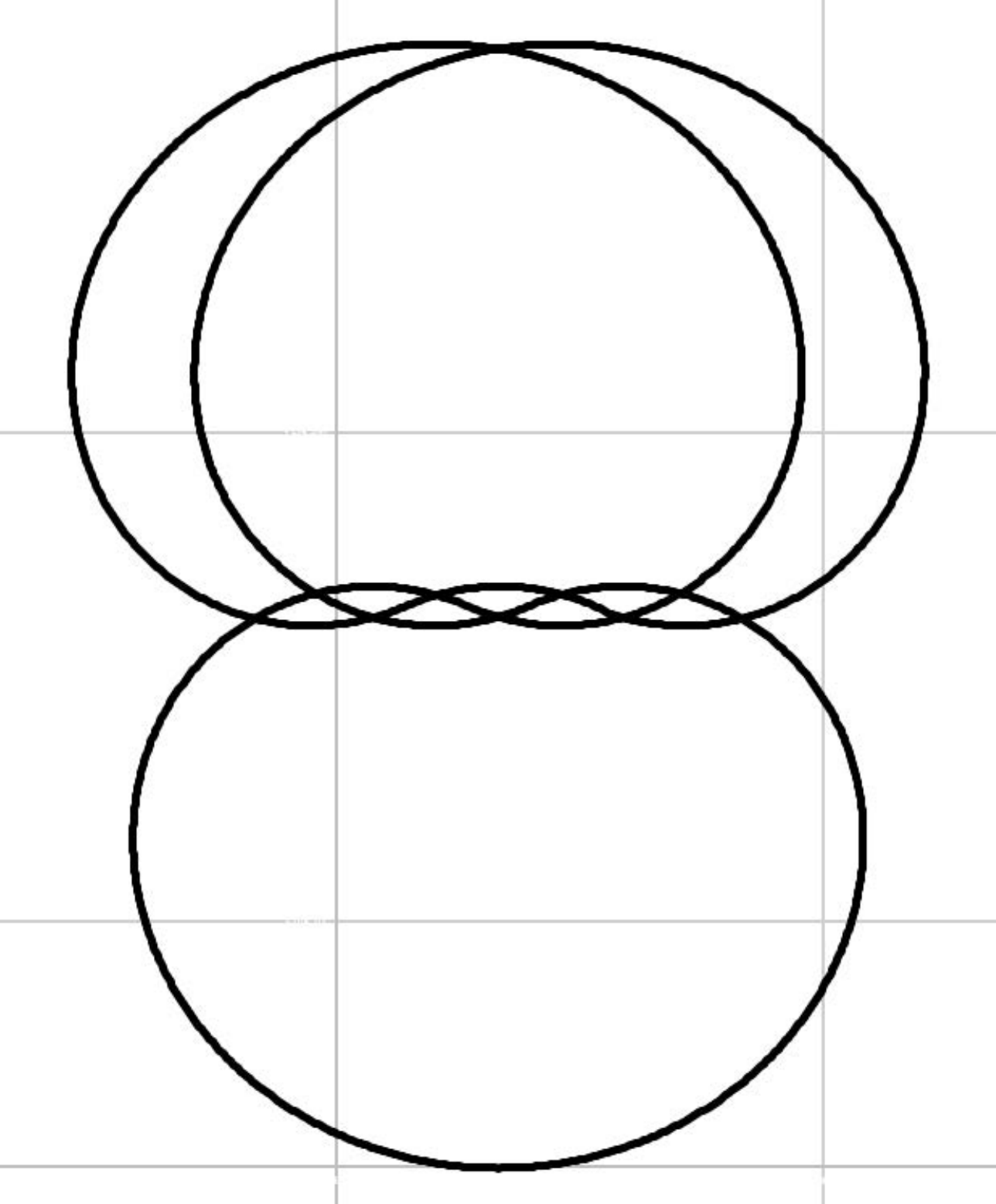}

(a)
\hskip 0.33\hsize 
(b)

\caption{Solutions of (\ref{4eq:gaugedMKdV2})
 of genus two, where $(k_1, k_2)$ are as follows:
(a): $(4.52,10.0)$, (b): $(7.47, 10.0)$.}
\label{4fg:lsoliton_g22}
\end{center}
\end{figure}

Lastly, we note that by similar computations, lifting 
 the condition CIII (\ref{eq:dvarphi_a}) but adding
CII $d s_{\ri}=0$, the gauged MKdV
equation  (\ref{4eq:gaugedMKdV2}) yields more intricate
shapes (cf. Figure~\ref{4fg:lsoliton_g22}).

\section{Conclusion}

We proposed an algebro-geometric method to plot 
loops, as trajectories of generalized of elastica.
We concluded that we cannot find the solution of the MKdV equation
(\ref{4eq:MKdV}) 
based on our algebraic data in Theorem \ref{4th:MKdVloop}
for $X_2$ $g=2$
and that  higher-genus curves ($g\ge 3$) are required 
to find the solution of (\ref{4eq:MKdV}).

Even though they are not solutions of the MKdV equation over $\RR$,
since they identically satisfy the MKdV equation over $\CC$ 
(\ref{4eq:loopMKdV2}) and are a natural generalization of Euler's elastica,
we demonstrated the typical shapes in Figures~\ref{4fg:lsoliton_g2}
and \ref{4fg:lsoliton_g22} in terms of the hyperelliptic functions
using the data in Theorem \ref{4th:MKdVloop}.
They partially recover some geometrical properties of the AFM or 
EMS images of the 
supercoiled DNA, cf., e.g., \cite{Betal,VV}.
In order to recover the shapes of the supercoiled DNA,
our algebro-geometrical approach is crucial.

By extension to higher genus curves and three dimensional space as
in \cite{C, GoldsteinLanger, Mat99a},
it is expected that the energy and topological properties of the 
supercoiled DNA would be clarified.

\bigskip

S.M. acknowledges support by JSPS KAKENHI Grant Number JP21K03289.
The authors thank the anonymous referees for crucial comments
and one of them for her/his suggestion to refer  to
\cite{Sa}.
%
%
%








\end{document}